# APPLICATION OF MODIFIED HYPERVIRIAL AND EHRENFEST THEOREMS AND SOME OF ITS CONSEQUENCES

ANZOR KHELASHVILI[1, 2,*] TEIMURAZ NADAREISHVILI[1,3]

[1] *Inst. of High Energy Physics, Iv. Javakhishvili Tbilisi State University, University Str. 9, 0109, Tbilisi, Georgia. E-mail:anzor.khelashvili@tsu.ge*

[2] *St.Andrea the First-called Georgian University of Patriarchate of Georgia, Chavchavadze Ave.53a, 0162, Tbilisi,Georgia.*

[3] *Faculty of Exact and Natural Sciences, Iv. Javakhishvili Tbilisi State University, Chavchavadze Ave 3, 0179, Tbilisi,Georgia.*

Corresponding author E-mail: *teimuraz.nadareishvili@tsu.ge*

**Abstract.**

It is well-known that owing to the restricted character of the area, in which the system is enclosed, additional "surface terms" emerge in the traditional form of hypervirial and/or Ehrenfest theorems. Especially, when one considers spherically symmetric potentials and operators the radial distance in spherical coordinates is restricted by a half-plane. Therefore the extra terms arise in this case as well as in view of boundary conditions at the origin of coordinates. We analyze the role of these terms for various model-potentials in the Schrodinger equation. We consider regular as well as soft-singular potentials and show that the inclusion of these terms is very essential in obtaining correct physical results. Among the well-known results some new ones are also derived.

# 1. INTRODUCTION

It is well known that when the system is located in finite volume the inclusion of boundary conditions becomes necessary as well as imposing restrictions on the allowed classes of wave functions. Therefore, it is remarkable to shed light whether or not some of well-known results are altered, when the problems connected to the boundary behavior come into play.

Remarkable contributions were made in this respect in series of papers [1-8], especially after the pioneering works of J.Esteve and collaborators [1, 9], where the strong mathematic definitions of operators and their various combinations are established according to their domains. It is also specified how the boundary contributions appear in various forms of virial-like relations. As regards of Ehrenfest-like theorems the strong mathematical grounds were considered in [10].

Such theorems are different forms of virial considerations, known already from classical mechanics. There are known many ways of their generalization in quantum mechanics. They mainly are rested on some manipulations upon the Hamiltonians and the Schrodinger equation. In derivation of such theorems, as a rule, the whole space had been considered [11,12].

In case of restricted motion wave functions and operators obey some boundary conditions and therefore several "surface terms" can be non-vanishing.

It is very interesting that in two and more dimensions, if we have a central symmetry, the radial variable is restricted by a half-space, and hence the boundary behavior can have an effect on corresponding theorems. Partly this problem was solved in [13], where elaboration of some fundamental relations in 3-dimensional quantum mechanics was made taking into account the restricted character of areas in radial distance. In such cases the boundary behavior of the radial wave function and singularity of operators at the origin of coordinates contribute to these relations. It was derived the relations between the average values of the operators' time derivative and the time derivative of average values of these operators, which is usually considered to be the same by definition [11]. The deviation from the known result was deduced and manifested by extra terms, depended on the boundary behavior mentioned above.

The general form for this extra term takes place in the hypervirial-like theorems. As a particular case, the virial theorem for Coulomb and oscillator potentials was considered and corrections to the Kramers' sum rule was derived. Moreover, the corrected Ehrenfest theorem was deduced and its consistency with real physical picture was demonstrated [13].

Our goal in the present paper is to apply this and related problems in 3-dimensional space, where some specific peculiarities occur, especially when the central symmetric potential in the Schrodinger equation is singular or considered operators are singular at the origin of coordinates.

Below we want to investigate the role of the extra contribution for various exactly solvable potential models in the Schrodinger equation.

This article is constructed as follows: In Sections 2 and 3 the brief review of theoretical reasoning is conveyed leading to modification of hypervirial and Ehrenfest theorems, in parallel their validity for Coulomb and oscillator potentials is verified. Remaining place is devoted to examination of the additional terms for other regular, as well as soft-singular potentials. We show that for all considered potentials the additional contribution works in correct direction and its presence in above mentioned theorems is essential.

## 2. MODIFIED HYPERVIRIAL THEOREM

### 2.1 General consideration

We consider the central symmetric potential $V(r)$ in the radial Schrodinger Hamiltonian

$$\hat{H} = \frac{\hbar^2}{2m}\left(-\frac{d^2}{dr^2} - \frac{2}{r}\frac{d}{dr}\right) + \frac{\hbar^2 l(l+1)}{2mr^2} + V(r) \qquad (2.1)$$

and explicitly time-independent operator $\hat{A}(r)$, which depends only on radial distance $r$. Under such circumstances the new hypervirial theorem for stationary states takes place [9,13]

$$\langle \phi, [\hat{H}, \hat{A}] \phi \rangle = i\hbar \Pi \qquad (2.2)$$

Moreover, we have derived there that the relation for time derivative of the operator's mean value has the form [13]

$$\frac{d\langle \hat{A} \rangle}{dt} = \frac{i}{\hbar}\langle [\hat{H}, \hat{A}] \rangle + \Pi \qquad (2.3)$$

In the last two relations the additional term is [13]

$$\Pi = i\frac{\hbar}{2m}\lim_{r \to 0}\left\{ r^2 \left[ \hat{A}R\frac{dR^*}{dr} - R^*\frac{d}{dr}(\hat{A}R) \right] \right\} \qquad (2.4)$$

Here $R(r)$ is the total radial wave function for bound state solution, which decreases at infinity, but, in general, gives a finite contribution at the origin of coordinates. For it we take

$$rR(r)\underset{r\to 0}{=}0 \qquad (2.5)$$

It corresponds to the Dirichlet boundary condition for the reduced wave function $u = rR$, and it follows under very general assumptions (for details, see [14-18]). Naturally, the degree of turning to zero explicitly depends on the potential under consideration. We follow to the traditional classification [13, 18]

- (1). Regular potentials:

$$\lim_{r \to 0} r^2 V(r) = 0 \qquad (2.6)$$

For which

$$R\underset{r\to 0}{=} C_1 r^l + C_2 r^{-(l+1)} \qquad (2.7)$$

Clearly, the second term is very singular and contradicts to (2.5) and therefore we must retain only the first term ($C_2 = 0$) or take

$$\lim_{r \to 0} R \approx C_1 r^l \tag{2.8}$$

- (2). <u>Singular potentials</u>, for which

$$\lim_{r \to 0} r^2 V(r) \to \pm\infty \tag{2.9}$$

For them the "falling to the center" happens and is not interesting for us now.

- (3). <u>"Soft singular" potentials</u>, for which

$$\lim_{r \to 0} r^2 V(r) \to \pm V_0, \quad (V_0 = const > 0) \tag{2.10}$$

Here the (+) sign corresponds to repulsion, while the (-) sign – to attraction. For such potentials the wave function has the following behavior [14-17]:

$$\lim_{r \to 0} R = a_{st} r^{-1/2+P} + a_{add} r^{-1/2-P} \equiv R_{st} + R_{add}, \tag{2.11}$$

where

$$P = \sqrt{(l+1/2)^2 - \frac{2mV_0}{\hbar^2}} \tag{2.12}$$

In the region $0 < P < 1/2$ the second solution satisfies also to the boundary condition (2.5), therefore it must be retained in general and hence the self-adjoint extension need to be performed [17]. As for the region $P \geq 1/2$ only the first (standard or regular) solution remains.

After this information let us return to consideration of the additional contribution (2.4). Below we restrict ourselves only by *regular solutions* both for regular and soft-singular potentials, i.e. the first terms in above equations, (2.8) and (2.11). It is obvious that upon calculation of the limit in Eq. (2.4) the behavior of the operator $\hat{A}$ in the origin will be also important. We take it as

$$\hat{A}(r) \sim \frac{1}{r^\beta}; \quad \beta > 0 \tag{2.13}$$

Here, it is implied not only explicit dependence on $r$, but also its scale dimension (derivative et al.). Taking all these into account, we chose

$$\hat{A} r^l \sim r^{l-\beta} \tag{2.14}$$

Then, we have (for regular potentials (2.6))

$$\Pi = \frac{i\hbar C_1^2}{2m} \lim_{r \to 0} r^2 \left\{ r^{l-\beta} l r^{l-1} - r^l \frac{d}{dr} r^{l-\beta} \right\} = \frac{i\hbar C_1^2}{2m} \lim_{r \to 0} r^{2l+1-\beta} \tag{2.15}$$

In order for this expression not to be diverging we must require

$$2l + 1 > \beta \tag{2.16}$$

In this case the additional term vanishes. If the inequality is reflected, then the divergent result will follow and we will be unable to write the equation (2.3), so in this case the time derivative is not defined on the whole.

On the other hand, if the operator is such that
$$2l+1=\beta \tag{2.17}$$
the extra term survives on the right-hand side (2.2)
$$\langle[\hat{H},\hat{A}]\rangle = -\frac{\hbar^2 C_1^2}{m}\left(l+\frac{1}{2}\right) \tag{2.18}$$
Just so happens also in case of the soft-singular potential (2.10), restricting ourselves by regular (or standard) solution only:
$$R \Rightarrow R_{st} = a_{st} r^{-1/2+P} \tag{2.19}$$
$$\Pi = \frac{i\hbar a_{st}^2}{2m}\lim_{r\to 0} r^2 \left\{ r^{-1/2+P-\beta}(-1/2+P)r^{P-3/2} - r^{-1/2+P}\frac{d}{dr}r^{-1/2+P-\beta}\right\} = i\hbar\frac{a_{st}^2 \beta}{2m}\lim_{r\to 0} r^{2P-\beta} \tag{2.20}$$
Here the finite contribution follows when
$$2P = \beta \tag{2.21}$$
Moreover, if $2P > \beta$, the additional contribution is zero.

In case of (2.21), we have
$$\langle[\hat{H},\hat{A}]\rangle = -\frac{\hbar^2 a_{st}^2}{m} P \tag{2.22}$$
The right sides of (2.18) and (2.22) coincide, when potential is regular, because for regular potential $P = l+\frac{1}{2}$.

### 2.2 Explicit form of hypervirial theorem for known operators. Verification for Coulomb and oscillator Potentials

Equations (2.18) and (2.22) are the final forms of hypervirial theorem in considered cases. For the further applications let us calculate their left-hand-sides for widespread used events.
We first discuss the consequences of these relations for specific radial operators and then move further.

Consider the following operator [19]
$$\hat{A} = \hat{p}_r f(r) \tag{2.23}$$
where $\hat{p}_r$ is a Hermitian operator of radial momentum
$$\hat{p}_r = \frac{\hbar}{i}\left(\frac{\partial}{\partial r}+\frac{1}{r}\right) \tag{2.24}$$
and $f(r)$ is a three-times differentiable function. Calculate the commutator
$$[\hat{H},\hat{A}] = -\frac{i\hbar}{2m}\left\{p_r^2 \frac{df}{dr}+\frac{df}{dr}p_r^2\right\} - \frac{\hbar^2}{2m}\frac{d^2 f}{dr^2}p_r + i\hbar f(r)\frac{dV}{dr} - i\hbar^3 \frac{l(l+1)}{mr^3}f \tag{2.25}$$

Entering here $\hat{p}_r^2$ and $\hat{p}_r$ can be rewritten by means of

$$-\frac{p_r^2}{2m} = -H + \frac{\hbar^2 l(l+1)}{2mr^2} + V \tag{2.26}$$

$$-\frac{\hbar}{m} i \frac{d^2 f}{dr^2} p_r = \left[H, \frac{df}{dr}\right] + \frac{\hbar^2}{2m} f''' \tag{2.27}$$

Finally,

$$[\hat{H}, \hat{A}] = i\hbar Q - \frac{3}{2} i\hbar [\hat{H}, f'] \tag{2.28}$$

where

$$Q = -2f'(H-V) + \hbar^2 \frac{l(l+1)}{m}\left[\frac{f'}{r^2} - \frac{f}{r^3}\right] - \frac{\hbar^2}{4m} f''' + f(r)V' \tag{2.29}$$

Then it follows

$$\langle Q \rangle - \frac{3}{2}\langle [\hat{H}, f']\rangle = \Pi \tag{2.30}$$

Here $\Pi$ is given by (2.15) and/or (2.20).

Taking into account (2.29) and (2.30) we obtain the most general hypervirial theorem for the Schrodinger equation [13]

$$\left\langle -2f'(H-V) + \hbar^2 \frac{l(l+1)}{m}\left[\frac{f'}{r^2} - \frac{f}{r^3}\right] - \frac{\hbar^2}{4m} f''' + f(r)V' \right\rangle =$$
$$= \frac{\hbar^2}{2m} \lim_{r \to 0}\left\{ r^2\left[ -R^2 f'' - \left(RR' + \frac{R^2}{r}\right)f' + \left(R'^2 - RR'' + \frac{R^2}{r^2}\right)f + \frac{3}{2} R^2 f''\right]\right\} \tag{2.31}$$

Now if one considers a particular case, $f = r^{S+1}$ applied in [20], a simple calculation gives:
- for regular potentials

$$\frac{2m}{\hbar^2}\left\{\left\langle r^{S+1} \frac{dV}{dr}\right\rangle + 2(S+1)\left[\langle r^S V\rangle - E\langle r^S\rangle\right]\right\} + \frac{1}{2} S\left[(2l+1)^2 - S^2\right]\langle r^{S-2}\rangle = (2l+1)^2 C_l^2 \delta_{S+1,-2l} \tag{2.32}$$

- and for soft-singular potentials

$$\frac{4m}{\hbar^2}\left\{\left\langle r^{S+1} \frac{dV}{dr}\right\rangle + 2(S+1)\left[\langle r^S V\rangle - E\langle r^S\rangle\right]\right\} + S\left[(2l+1)^2 - S^2\right]\langle r^{S-2}\rangle = a_{st}^2 S(S-2P)\delta_{2P,-S} \tag{2.33}$$

Analogous to relation (2.32) was derived in [21] by different method. We display it here in order to show that our consideration gives the correct result. Note that if we substitute $s = j-1$ into the Eq. (2.32), we obtain the relation (11) of J.Sukumar [21]. See also [22]. More details for a generalization of Quigg and Rosner method [23] in connection to [21] may be found in [22]. Therefore, it means that above calculation by commutator gives the same result as a calculation based on manipulations applied in [21-23].

As for (2.33), it is really a new relation, which incorporates a soft-singular potential too. Using it for the Coulomb and oscillator potentials, we have, correspondingly:

$$2E(s+1)\langle r^s \rangle + e^2(2s+1)\langle r^{s-1} \rangle + \frac{s\hbar^2}{4m}\left[s^2 - (2l+1)^2\right]\langle r^{s-2} \rangle = -\frac{\hbar^2}{2m}(2l+1)^2 C_l^2 \delta_{s+1,-2l} \quad (2.34)$$

and

$$2E(s+1)\langle r^s \rangle - m\omega^2(s+2)\langle r^{s+2} \rangle + \frac{s\hbar^2}{4m}\left[s^2 - (2l+1)^2\right]\langle r^{s-2} \rangle = -\frac{\hbar^2}{2m}(2l+1)^2 C_l^2 \delta_{s+1,-2l} \quad (2.35)$$

If one compares these relations with the corresponding relations for the Coulomb potential $V = -\frac{e^2}{r}$ and oscillator $V = \frac{m}{2}\omega^2 r^2$ as for regular potentials [19,24-25], one obtains the difference between these two in the non-zero right-hand-side contributions. Exactly these sides balance obtained sum rules, as we'll see below.

For example, modified Kramers' relation according to Eq.(2.34) for the Coulomb potential, looks like:

$$2E(s+1)\langle r^s \rangle + e^2(2s+1)\langle r^{s-1} \rangle + \frac{s\hbar^2}{4m}\left[s^2 - (2l+1)^2\right]\langle r^{s-2} \rangle = -\frac{\hbar^2}{2m}(2l+1)^2 \tilde{C}_{nl}^2 \left(\frac{2}{na_0}\right)^{2l} \delta_{s+1,-2l} \quad (2.36)$$

On the other hand, it follows from the Coulomb wave function

$$R_{nl}(r) = \tilde{C}_{nl}\left(\frac{B}{n}\right)^l r^l e^{-\frac{Br}{2n}} F\left(-n+l+1, 2l+2, \frac{Br}{n}\right), \quad \tilde{C}_{nl} = \frac{B}{n^2(2l+1)!}\left(\frac{B}{2}\right)^{\frac{1}{2}}\sqrt{\frac{(n+l)!}{(n-l-1)!}}, \quad (2.37)$$

that

$$C_l = \tilde{C}_{nl}\left(\frac{B}{n}\right)^l, \quad \frac{B}{n} = \frac{2}{na_0}, \quad \text{and} \quad a_0 = \frac{\hbar^2}{me^2} \quad (2.38)$$

Then the Kramers' modified relation (2.36) takes the form

$$4El\langle r^{-2l-1} \rangle + e^2(4l+1)\langle r^{-2l-2} \rangle = \frac{\hbar^2}{2m}(2l+1)^2 \tilde{C}_{nl}^2 \left(\frac{2}{na_0}\right)^{2l} \quad (2.39)$$

For verification of its validity, consider some of first values of $l$:

(i)      $l = 0$.

This case corresponds to $s = -1$, i.e. $\hat{A} = \hat{p}_r$. Then Eq. (2.36) gives

$$e^2\left\langle\frac{1}{r^2}\right\rangle = \frac{\hbar^2}{2m}C_{n0}^2, \tag{2.40}$$

whereas the ordinary Kramers' formula gives a contradictory result - $\langle r^{-2}\rangle = 0$. The Kramers' relation does not involve the s-wave $l = 0$. So, our theorem is complimentary to the Kramers' one.

If one considers other waves $l > 0$ then the Kramers' formula will be successful, because in this case the additional contribution vanishes, as the Coulomb potential is a regular one.

The analogous situation appears in oscillator potential case. For s-wave the additional term straightens out the validity of Kramers' relation. For details see [13].

## 3. MODIFIED EHRENFEST THEOREM

The Ehrenfest's equations signify that the average values of position and linear momentum operators evolve classically. We now simply analyze what happens with the Ehrenfest theorem in ordinary quantum mechanics in light of influences of presented boundary behavior in spherical coordinates.

Consider again the operator of radial momentum $\hat{p}_r$ (2.24) and substitute it into Eq. (2.3), we have

$$\frac{d\langle\hat{p}_r\rangle}{dt} = \frac{i}{\hbar}\langle[\hat{H},\hat{p}_r]\rangle + \Pi \tag{3.1}$$

where

$$\Pi = \frac{a_{st}^2\hbar^2}{2m}\lim_{r\to 0} r^2\left\{\left(\frac{\partial}{\partial r}+\frac{1}{r}\right)\left[r^{-1/2+P}\right]\left(-\frac{1}{2}+P\right)\left(r^{-3/2+P}\right) - \left(r^{-1/2+P}\right)\frac{d}{dr}A\left(\frac{\partial}{\partial r}+\frac{1}{r}\right)\left[r^{-1/2+P}\right]\right\} =$$

$$= \frac{a_{st}^2\hbar^2}{2m}\left(\frac{1}{2}+P\right)\lim_{r\to 0} r^{2P-1} \tag{3.2}$$

It is clear from this relation that for $2P > 1$, $\Pi = 0$, while for $2P < 1$, it diverges. But for $2P = 1$ it survives

$$\Pi = \frac{a_{st}^2\hbar^2}{2m} \tag{3.3}$$

Therefore, for singular potential the usual Ehrenfest theorem is applicable only in the first case $2P > 1$. In other cases, the additional term (3.3) appears or theorem does not exist at all. Remember that *in traditional textbooks this fact is not mentioned.*

Let us now calculate the commutator in (3.1). We find

$$[\hat{H},\hat{p}_r] = \frac{\hbar^2 l(l+1)}{2m}\left[\frac{1}{r^2},p_r\right] + [V(r),p_r] \tag{3.4}$$

But

$$\left[\frac{1}{r^2}, p_r\right] = -i\frac{2\hbar}{r^3}; \quad [V, p_r] = i\hbar\frac{\partial V}{\partial r} = -i\hbar F_r \tag{3.5}$$

Where $F_r$ is a "radial force". Therefore, we get

$$\frac{d\langle\hat{p}_r\rangle}{dt} = \frac{\hbar^2 l(l+1)}{m}\left\langle\frac{1}{r^3}\right\rangle + \langle F_r\rangle + \Pi \tag{3.6}$$

It is the modified Ehrenfest theorem for time evolution of radial momentum (Newton's "second law). *This relation is a new one.* For regular potentials, when $P = l+1/2$, as it seems from (3.2), only in case $l > 0$ it follows $\Pi = 0$. As for $l = 0$, we have

$$\Pi = \frac{C_1^2 \hbar^2}{2m}, \text{ evidently } \quad C_1 = C_0 = a_{st} \tag{3.7}$$

We conclude here that *for regular potentials the usual Ehrenfest theorem is valid only in case $l > 0$ while in case $l = 0$, there appears an extra term* (3.7).

Now let us show that Eq. (3.6) gives correct results for Coulomb potential. Note that for stationary states time derivative of mean value of time independent operator must be zero. Therefore, the left-hand side of (3.8) is zero and it remains

$$\frac{\hbar^2 l(l+1)}{m}\left\langle\frac{1}{r^3}\right\rangle + \langle F_r\rangle + \Pi = 0 \tag{3.8}$$

Let us check this relation. Consider first the case of nonzero angular momentum $l > 0$. In this case, $\Pi = 0$ and remaining terms on the right-hand side $\frac{\hbar^2 l(l+1)}{m}\left\langle\frac{1}{r^3}\right\rangle + \langle F_r\rangle$ compensate each other's for Coulomb potential [26-27].

On the other hand, the case $l = 0$ is more interesting and crucial. In this case we have no centrifugal term, and the additional term is given by (3.7),

$$0 = \langle F_r\rangle + \frac{C_1^2 \hbar^2}{2m} \tag{3.9}$$

But for Coulomb potential

$$C_1 = \frac{4}{n^3 a_0^3} \quad \text{and} \quad \langle F_r\rangle = \left\langle -\frac{e^2}{r^2}\right\rangle, \tag{3.10}$$

Moreover

$$\left\langle\frac{1}{r^2}\right\rangle = \frac{1}{n^3 a_0^2}\frac{1}{(l+1/2)} \rightarrow \frac{2}{n^3 a_0^2} \tag{3.11}$$

So, we should have

$$\frac{C_1^2 \hbar^2}{2m} = \frac{2e^2}{n^3 a_0^2} \tag{3.12}$$

It is easy to verify that this equality is correct, because together with Eq. (3.10) it gives the correct value for the Bohr's first orbit radius $a_0 = \frac{\hbar^2}{me^2}$.

On the other hand, without the additional term a meaningless result follows: $\langle r^{-2} \rangle = 0$.

All the above consideration tells us that the inclusion of the additional term into virial and Ehrenfest theorems is necessary.

## 4. OTHER REGULAR POTENTIALS

Let us now consider other regular potentials which have a wide application in quantum mechanics. Our aim will be further analyzing the modified Ehrenfest theorem (3.6) first for various regular potentials, and then for singular ones also.

Remember that for regular potentials in case of nonzero angular momentum, $l > 0$ the surface term is absent, $\Pi = 0$. Moreover for stationary case, the left-hand side of Eq. (3.6) must be zero, i.e. the following sum rule must be valid

$$\langle F_r \rangle = -\frac{\hbar^2 l(l+1)}{m} \left\langle \frac{1}{r^3} \right\rangle \tag{4.1}$$

Because $F_r = -\frac{dV}{dr}$, (4.1) takes more familiar form

$$\left\langle \frac{dV}{dr} \right\rangle = \frac{\hbar^2 l(l+1)}{m} \left\langle \frac{1}{r^3} \right\rangle \tag{4.2}$$

Remember that this relation can be derived from Eq. (2.31) if we take $s = -1$, $l > 0$. (4.2) is a well-known formula in current literature, see [23,28]. We have proved it from modified Ehrenfest equation (3.6).

This relation has to be valid for any regular potentials, therefore it can be considered as a new "virial" relation, because it relates various average values.

Let us consider the particular class of regular potentials

$$V = V_0 r^k; \qquad k > -2 \tag{4.3}$$

Then (4.2) gives

$$kV_0 \cdot \langle r^{k-1} \rangle = \frac{\hbar^2 l(l+1)}{m} \left\langle \frac{1}{r^3} \right\rangle \tag{4.4}$$

For example, oscillator potential

$$V = V_0 r^2 \tag{4.5}$$

corresponds to

$$2V_0 \cdot \langle r \rangle = \frac{\hbar^2 l(l+1)}{m} \left\langle \frac{1}{r^3} \right\rangle \tag{4.6}$$

One can verify this relation for the several first states. Radial wave functions for $1s$ and $1p$ states are [11]

$$R_{00} = 2\sqrt{\frac{\alpha^3}{\pi}} e^{-\alpha r^2/2} \equiv C_{00} e^{-\alpha r^2/2}; \quad R_{01}(r) = \sqrt{\frac{8}{3}} \frac{\alpha^{5/4}}{\pi^{1/4}} r e^{-\alpha r^2/2} \equiv C_{01} r e^{-\alpha r^2/2}; \quad \alpha = \frac{m\omega}{\hbar} \tag{4.7}$$

In general, for averaging values by means of radial functions we have the following relation

$$\langle r^k \rangle = \int_0^\infty R_{n,l}^2(r) r^{k+2} dr \tag{4.8}$$

We need above the following integrals Eq. [(4.8) with $k = 1$ and $k = -3$]. Using [29]

$$\int_0^\infty x^{2n+1} e^{-px^2} dx = \frac{n!}{2 p^{n+1}}; \qquad p > 0 \tag{4.9}$$

we obtain

$$\langle r \rangle = C_{01}^2 \frac{1}{\alpha^3}, \qquad \left\langle \frac{1}{r^3} \right\rangle = C_{01}^2 \frac{1}{2\alpha} \tag{4.10}$$

Inserting these results into Eq. (4.4) for $l = 1$ we get the relation

$$2V_0 \frac{1}{\alpha^3} = \frac{2\hbar^2}{m} \frac{1}{2\alpha} \tag{4.11}$$

On the other hand, from (4.3) with $k = 2$ and (4.5), $\alpha = \frac{m\omega}{\hbar}$ (4.7) it follows

$$V_0 = \frac{\alpha \omega \hbar}{2} \tag{4.12}$$

Substituting it into (4.11) we are convinced that this relation is satisfied exactly or the relation (4.6) is correct for $l = 1$. Correctness of Eq. (4.6) may be verified easily for all values of $l$ applying the so-called Pasternak-type inversion property [30] (here the dimensionless values are used):

$$\langle r^{-p-2} \rangle = \frac{\Gamma(K+(n-p)/2-1)}{\Gamma(K+(n+p)/2)} \langle r^p \rangle \qquad \text{(for all convergent integrals)} \tag{4.13}$$

The form (4.4) has an independent significance, as it connects different mean values. It can be very useful even in cases when the Schrodinger equation is not solvable analytically. For example, in case of linear potential

$$V = V_0 r \qquad (4.14)$$

The Schrodinger equation is solvable only in S-state $l = 0$. But the above (4.4) formula for (4.14) potential gives for any $l > 0$

$$\left\langle \frac{1}{r^3} \right\rangle = \frac{mV_0}{\hbar^2 l(l+1)} \qquad (4.15)$$

Let us consider the well-known *quarkonium potential* [23]

$$V(r) = -\frac{\alpha}{r} + V_0 r \qquad (4.16)$$

It follows from (4.4) that

$$V_0 + \alpha \left\langle \frac{1}{r^2} \right\rangle = \frac{\hbar^2 l(l+1)}{m} \left\langle \frac{1}{r^3} \right\rangle \qquad (4.17)$$

Here we do not know the exact solution of Schrodinger equation, but different averages are related.

The case $l = 0$ is more interesting, because the additional term contributes. Now it follows from (3.6) and (3.7) that for stationary states

$$\frac{C_1^2 \hbar^2}{2m} = \left\langle \frac{dV}{dr} \right\rangle \qquad (4.18)$$

Remark here that $C_1 = R_0(0) = \sqrt{4\pi} \psi_{00}(0)$, but because $Y_{00}(\theta, \varphi) = \frac{1}{\sqrt{4\pi}}$, then the total wave function is

$$\psi_{00} = \frac{R_0}{\sqrt{4\pi}} \qquad (4.19)$$

Therefore (4.18) gives the well-known relation [23,28]

$$|\psi_{00}(0)|^2 = \frac{m}{2\pi\hbar^2} \left\langle \frac{dV}{dr} \right\rangle \qquad (4.20)$$

This shows ones again that the inclusion of the additional term is necessary. The ordinary Kramers' theorem does not work in this case.

It is easy exercise to verify (4.18) for linear potential (4.14) because solution in Airy functions is well known [11]. The reduced Schrodinger equation for this problem has a form

$$u'' + xu = 0 \qquad (4.21)$$

where

$$u = rR; \quad x = \left(\frac{2mV_0}{\hbar^2}\right)^{\frac{1}{3}}\left(\frac{E_{n_r}}{V_0} - r\right) \qquad (4.22)$$

Its solution that falls at infinity is an Airy function

$$u_{n_r}(r) = N_{n_r} Ai(-\xi) = NAi\left[\left(\frac{2mV_0}{\hbar^2}\right)^{\frac{1}{3}}\left(r - \frac{E_{n_r}}{V_0}\right)\right] \qquad (4.23)$$

$n_r$ - is a radial quantum number describing excitation states. Energy levels can be found from zero boundary condition $u_{n_r}(0) = 0$

$$Ai\left[-\left(\frac{2mV_0}{\hbar^2}\right)^{\frac{1}{3}}\frac{E_{n_r}}{V_0}\right] = 0 \qquad (4.24)$$

Taking this into account we derive

$$\lim_{r \to 0} u_{n_r}(r) = N_{n_r} \lim_{r \to 0} Ai\left[\left(\frac{2mV_0}{\hbar^2}\right)^{\frac{1}{3}}\left(r - \frac{E_{n_r}}{V_0}\right)\right] =$$
$$= N_{n_r} \left(\frac{2mV_0}{\hbar^2}\right)^{\frac{1}{3}} Ai'\left(-\left(\frac{2mV_0}{\hbar^2}\right)^{\frac{1}{3}}\frac{E_{n_r}}{V_0}\right) r \qquad (4.25)$$

Therefore from (2.8)

$$C_{1n_r} = N_{n_r} \left(\frac{2mV_0}{\hbar^2}\right)^{\frac{1}{3}} Ai'\left(-\left(\frac{2mV_0}{\hbar^2}\right)^{\frac{1}{3}}\frac{E_{n_r}}{V_0}\right) \qquad (4.26)$$

where eigenvalues $E_{n_r}$ are determined from (4.24) by zeros of the Airy function

$$E_{n_r} = \left(\frac{V_0^2 \hbar^2}{2m}\right)^{\frac{1}{3}} \alpha_{n_r+1}; \qquad n_r = 0, 1, 2... \qquad (4.27)$$

Here $\alpha_{n_r}$ are placed in increasing order. Then after substituting all of that into Eq. (4.18), which has the following form in this case

$$\frac{C_{1n_r}^2 \hbar^2}{2m} = V_0 \qquad (4.28)$$

we derive

$$N_{n_r}^2 \left(\frac{2mV_0}{\hbar^2}\right)^{\frac{2}{3}} Ai'^2\left(\alpha_{n_r+1}\right)\frac{\hbar^2}{2m} = V_0 \qquad (4.29)$$

From which we obtain the normalization constant

$$N_{n_r} = \left(\frac{2mV_0}{\hbar^2}\right)^{\frac{1}{6}} \frac{1}{Ai'\left(\alpha_{n_r+1}\right)} \qquad (4.30)$$

It coincides with the results, obtained in the book [31].

Linear and simple harmonic potentials are considered previously by C.Sukumar [21]. It is easy to convince that any powers of radial distance can be found by successive application of (2.32). So, this relation is a successive recurrence formula for power low regular potentials.

### 4.1. Other solvable potentials in $l=0$ state

It may be remarked that for $l=0$ states the Schrodinger equation has solution for several potentials, which have a wide application in physics. Note that such examples are not considered in [21].

Below we consider such problems:

1. *Exponential potential,*

$$V = -V_0 e^{-\frac{r}{a}} \qquad (4.31)$$

This potential is well-known from deuterium problem in nuclear physics.

The wave function is [32]

$$u = rR = C_{0n_r} J_p\left(\lambda e^{-\frac{r}{2a}}\right), \qquad (4.32)$$

where $C_{0n_r}$ is a normalization constant, index zero means that $l=0$. $J_p$ is the Bessel function of order $p$ and

$$p = 2\sqrt{\frac{-2mE_{n_r 0}}{\hbar^2}} a; \qquad \lambda = 2\sqrt{\frac{8mV_0}{\hbar^2}} a \qquad (4.33)$$

Energy spectrum is obtained from the condition

$$u(0) = J_p(\lambda) = 0 \qquad (4.34)$$

But the mean values are not known. In our case (4.18) gives

$$\left(C_{0n_r}\right)^2 \frac{V_0}{a} \int_0^\infty J_p^2\left(\lambda e^{-\frac{r}{2a}}\right) e^{-\frac{r}{a}} dr = \frac{C_1^2 \hbar^2}{2m} \qquad (4.35)$$

and here $C_1$ is to be established. Evidently

$$\lim_{r \to 0} u(r) = C_{0n_r} \lim_{r \to 0} J_p\left(\lambda e^{-\frac{r}{2a}}\right) \approx C_{0n_r} \lim_{r \to 0} J_p\left(\lambda - \lambda \frac{r}{2a}\right) = C_{0n_r}\left\{J_p(\lambda) - J_p'(\lambda)\frac{\lambda r}{2a}\right\} =$$

$$= -C_{0n_r} J_p'(\lambda) \frac{\lambda r}{2a} \qquad (4.36)$$

Therefore

$$C_1 = -C_{0n_r} J_p'(\lambda) \frac{\lambda}{2a} \qquad (4.37)$$

This relation can be obtained also more simply from the definition (2.7)

$$C_1 = R(0) = u'(0) \qquad (4.38)$$

Substituting (4.37) into (4.35) we get

$$\left(C_{0n_r}\right)^2 \frac{V_0}{a} \int_0^\infty J_p^2\left(\lambda e^{-\frac{r}{2a}}\right) e^{-\frac{r}{a}} dr = (C_{0n_r})^2 \left[J_p'(\lambda)\right]^2 \frac{\lambda^2}{4a^2} \frac{\hbar^2}{2m} \qquad (4.39)$$

or

$$\int_0^\infty J_p^2\left(\lambda e^{-\frac{r}{2a}}\right) e^{-\frac{r}{a}} dr = \left[J_p'(\lambda)\right]^2 \frac{\lambda^2}{a} \frac{\hbar^2}{8mV_0} \qquad (4.40)$$

Taking into account notations (4.33) it follows the resulting integral

$$\int_0^\infty J_p^2\left(\lambda e^{-\frac{r}{2a}}\right) e^{-\frac{r}{a}} dr = a\left[J_p'(\lambda)\right]^2 \qquad (4.41)$$

The only unknown parameter here is $\lambda$, but really it is obtainable from zeros of Bessel function (See, Eq. (4.34)). Our above derivation is much simpler than that given in [33].

Using (2.31) one can derive other helpful integral also: substituting there $s = 0, l = 0$ the virial theorem follows in the form

$$\left\langle r\frac{dV}{dr}\right\rangle = 2\bigl[E-\langle V\rangle\bigr] \tag{4.42}$$

Explicitly it gives

$$\frac{V_0}{a}\left\langle re^{-r/a}\right\rangle = 2E + 2V_0\left\langle e^{-r/a}\right\rangle \tag{4.43}$$

It means

$$\frac{V_0}{a}\int_0^\infty J_p^2\!\left(\lambda e^{-\frac{r}{2a}}\right)e^{-\frac{r}{a}}r\,dr = 2E + 2V_0\int_0^\infty J_p^2\!\left(\lambda e^{-\frac{r}{2a}}\right)e^{-\frac{r}{a}}dr \tag{4.44}$$

So, this new integral is expressed by previous one (4.41). According to notations (4.33), it can be reduced to the alternate form

$$\int_0^\infty J_p^2\!\left(\lambda e^{-\frac{r}{2a}}\right)e^{-\frac{r}{a}}r\,dr = -\frac{2ap^2}{\lambda^2} + 2a^2\bigl[J_p'(\lambda)\bigr]^2 \tag{4.45}$$

Other solvable cases are done in the Appendix below.

## 5. HYPERVIRIAL THEOREM FOR THE SOFT-SINGULAR POTENTIALS

As was noted in the introduction, our main goal in this article is a verification of validity of a new modified virial theorem for various potentials, included the soft-singular ones. For such potentials the master uquation is (2.33). This equation involves several parameters. Assigning them specific values, one can derive various sum rules for a given potential. We do not report them here. Because a nontrivial factor is an extra term, we concentrate our attention to cases, when this term is not zero, i.e. when the right-hand side of (2.33) is nonvanishing. It happens when

$$s = -2P \tag{5.1}$$

In this case (2.33) takes the following form

$$\frac{4m}{\hbar^2}\left\{\left\langle r^{1-2P}\frac{dV}{dr}\right\rangle + 2(1-2P)\bigl\langle r^{-2P}(V-E)\bigr\rangle\right\} - 2P\bigl[(2l+1)^2 - 4P^2\bigr]\bigl\langle r^{-2P-2}\bigr\rangle = 8P^2 a_{st}^2 \tag{5.2}$$

Let us introduce a new parameter k, by the relation

$$2P = 2k+1 \tag{5.3}$$

It coincides to the power degree of the regular part of (2.11), $k = -1/2 + P$. It follows from (5.2)

$$(2k+1)^2 \left|R_{n,l}^{(k)}(0)\right|^2 = (k!)^2 \frac{2m}{\hbar^2}\left\{\left\langle r^{-2k}\frac{dV}{dr}\right\rangle + 4k\left\langle r^{-2k-1}(E-V)\right\rangle\right\} - (k!)^2(2k+1)\frac{4mV_0}{\hbar^2}\left\langle r^{-2k-3}\right\rangle \quad (5.4)$$

Here $R_{n,l}^{(k)}(0)$ denotes $k^{th}$ derivative and the evident relation is used

$$a_{st} = \frac{R_{n,l}^{(k)}(0)}{k!} \quad (5.5)$$

The equation (5.4) is a generalization of Khare's known relation[34]. Indeed, when $V_0 = 0$ it follows Khare's relation in case of regular potentials

$$(2l+1)^2 \left|R_{n,l}^{(l)}(0)\right|^2 = \frac{2m}{\hbar^2}(l!)^2\left[\left\langle \frac{1}{r^{2l}}\frac{dV}{dr}\right\rangle + 4l\left\langle \frac{E-V}{r^{2l+1}}\right\rangle\right] \quad (5.6)$$

Morever the form (5.4) may be used in the Van Royen-Weisskopf formula [35] for decay rates in case of soft-singular potential.

In course of deriving these equations we have used the definition (2.12), from which it follows

$$l(l+1) - \frac{2mV_0}{\hbar^2} = k(k+1), \qquad l > k \quad (5.7)$$

Let us now study the following class of soft-singular potentials

$$V(r) = -\frac{V_0}{r^2} + U(r) \quad (5.8)$$

Where $U(r)$ is less singular, than $r^{-2}$. Evidently eq (5.4) is legitimated for such potentials. It follows

$$(2k+1)^2 \left|R_{n,l}^{(k)}(0)\right|^2 = \frac{2m}{\hbar^2}(k!)^2\left[\left\langle \frac{1}{r^{2k}}\frac{dU}{dr}\right\rangle + 4k\left\langle \frac{E-U}{r^{2k+1}}\right\rangle\right] \quad (5.9)$$

i.e. formally it coincides to (5.6), but is valid for all $k < l$. One can easily repeat all theorems, given in [34] by suitable changes $V \to U$, $l \to k$.

Substitute (5.8) into (2.33), we arrive at

$$\frac{4m}{\hbar^2}\left\{\left\langle r^{s+1}\frac{dU}{dr}\right\rangle + 2(s+1)\left[\left\langle r^s(U-E)\right\rangle\right]\right\} + \left\{s\left[(2l+1)^2 - s^2 - \frac{8mV_0}{\hbar^2}\right]\right\}\left\langle r^{s-2}\right\rangle = a_{st}^2 s(s-2P)\delta_{2P,-s} \quad (5.10)$$

But when $s = -2P$ and because $(l+1/2)^2 - \frac{2mV_0}{\hbar^2} = P^2$, this equation simplifies significantly

$$\frac{4m}{\hbar^2}\left\{\left\langle r^{1-2P}\frac{dU}{dr}\right\rangle + 2(1-2P)\left\langle r^{-2P}(U-E)\right\rangle\right\} = 8P^2 a_{st}^2 \tag{5.11}$$

We see, that the trace of singular part is eliminated completely apart from the exponent of wave function behavior at the origin $P$ and from the solution itself in the averaging procedure.

If now we take the potential like (5.8), it follows

$$a_{st}^2 = \frac{mg}{2\hbar^2 P^2}(\beta - 4P + 2)\left\langle r^{\beta-2P}\right\rangle - \frac{m}{2\hbar^2 P^2}2(1-2P)E\left\langle r^{-2P}\right\rangle \tag{5.12}$$

Because a parameter $P$ is arbitrary so long, consider case, when the first term here vanishes $2P = \frac{\beta}{2}+1$. Then (5.12) reads as

$$(\beta+2)^2 a_{st}^2 = \frac{8m\beta E}{\hbar^2}\left\langle r^{-\beta/2-1}\right\rangle \tag{5.13}$$

Let us study which values of $\beta$ are allowed in case of $l=0$ and $l \neq 0$ here. From the definition (2.12) it follows

$$l(l+1) - \frac{8mV_0}{\hbar^2} = \frac{\beta^2}{4} + \beta \tag{5.14}$$

And it can be easily verified that this definition does not contradict to the physically interesting potentials, namely: $\beta = -1$ (valence electron model) and $\beta = 2$ (singular oscillator, with $l \geq 2$), considering of which we are going to.

## 6. THE VALENCE ELECTRON AND THE SINGULAR OSCILLATOR MODELS

Consider now the valence electron model potential

$$V = \frac{-V_0}{r^2} - \frac{\alpha}{r} \tag{6.1}$$

Such potential arises in atomic physics in describing of alkali metals [11]. We substitute into (5.12) $\beta = -1, \quad g = -\alpha$. It follows

$$a_{st}^2 = -\frac{8mE}{\hbar^2}\left\langle r^{-1/2}\right\rangle \tag{6.2}$$

We are going to check this formula. The standard solution in case of potential (6.1) is [11,16]

$$R = C_1 \rho^{-1/2+P} e^{-\rho/2} F(1/2 + P - \lambda, 1 + 2P; \rho) \tag{6.3}$$

where

$$\rho = \sqrt{\frac{-8mE}{\hbar^2}} r \equiv \eta r; \quad \eta = \left(-\frac{8mE}{\hbar^2}\right)^{\frac{1}{2}}; \quad \lambda = \frac{2m\alpha}{\hbar\sqrt{-8mE}} > 0; \quad E < 0 \tag{6.4}$$

Considering the behavior of (6.3) at the origin, one derives

$$a_{st} = C_1 \eta^{P-1/2} \tag{6.5}$$

On the other hand, together with Eq. (6.2) and $2P = 1/2$, we have to verify the relation

$$\langle r^{-1/2} \rangle = -\frac{\hbar^2}{8mE} a_{st}^2 = C_1^2 \eta^{-2} \eta^{-1/2} = C_1^2 \eta^{-5/2} \tag{6.6}$$

The calculation of averaged values of arbitrary powers can be performed by using very powerful integral, given in the book [11]:

$$\langle r^s \rangle = \int_0^\infty e^{-kz} z^{\nu-1} \left[F(-n,\gamma,kz)\right]^2 dz = \frac{\Gamma(\nu) n!}{k^\nu \gamma(\gamma+1)\dots(\gamma+n-1)} \times$$
$$\times \left[1 + \sum_{s=0}^{n-1} \frac{n(n-1)\dots(n-s)(\gamma-\nu-s-1)(\gamma-\nu-s)\dots(\gamma-\nu+s)}{\left[(s+1)!\right]^2 \gamma(\gamma+1)\dots(\gamma+s)}\right] \tag{6.7}$$

Taken into account that in our case

$$k = \eta; \quad \nu = 2P + 2 + s; \quad \gamma = 2P + 1; \quad 1/2 + P - \lambda = -n_r, \quad n_r = 0,1,2,\dots \tag{6.8}$$

and $a(a+1)\dots(a+n-1) = \dfrac{\Gamma(a+n)}{\Gamma(a)}$, we obtain the general formula

$$\langle r^s \rangle = \frac{C_1^2 \Gamma(2P+2+s)\Gamma(2P+1) n_r!}{\eta^{s+3} \Gamma(2P+n_r+1)} \times$$
$$\times \{1 + \frac{n_r(2+s)(1+s)}{1^2(2P+1)} + \frac{n_r(n_r-1)(3+s)(2+s)(1+s)s}{1^2 2^2 (2P+1)(2P+2)} + \dots \tag{6.9}$$
$$+ \frac{n_r!(-1-s-n_r)\dots(-1-s+n_r-1)}{1^2 2^2 \dots (2P+1)(2P+2)\dots(2P+n_r)}\}$$

Substituting $s = 0$ into (6.9), it follows the expression for the normalization constant

$$C_1^2 = \frac{\eta^3}{n_r!(2n_r + 2P + 1)} \frac{\Gamma(2P+1+n_r)}{\Gamma^2(2P+1)} \tag{6.10}$$

For some negative integer values of $s$ the Eq. (6.9) gives also closed answers. For example,

$$(s=-1) \quad \rightarrow \quad \left\langle \frac{1}{r} \right\rangle = \frac{C_1^2 \Gamma^2 (2P+1) n_r!}{\eta^2 \Gamma(2P+n_r+1)} \tag{6.11}$$

$$(s=-2) \quad \rightarrow \quad \left\langle \frac{1}{r^2} \right\rangle = \frac{C_1^2 \Gamma^2 (2P+2) n_r!}{\eta \Gamma(2P+n_r+1) 2P} \tag{6.12}$$

Moreover, for positive degrees

$$(s=1) \quad \rightarrow \quad \langle r \rangle = \frac{C_1^2 \Gamma(2P+3)\Gamma(2P+1) n_r!}{\eta^{s+3} \Gamma(2P+n_r+1)} \times$$

$$\times \left\{ 1 + \frac{n_r 3 \cdot 2}{(2P+1)} + \frac{n_r (n_r-1) 4 \cdot 3 \cdot 2}{1^2 2^2 (2P+1)(2P+2)} \right\} \tag{6.13}$$

etc.

It is evident that for integer $s$ (positive or negative) the series (6.9) would truncate somewhere. Consider now fractional values of $s$. In such cases it is more convenient to consider some of first values of $n_r$. For example for $s = -1/2$

$$(n_r = 0), \quad \langle r^{-1/2} \rangle = \frac{C_1^2 \Gamma(2P+3/2)}{\eta^{5/2}}, \tag{6.14}$$

$$(n_r = 1), \quad \langle r^{-1/2} \rangle = \frac{C_1^2 \Gamma(2P+3/2)}{\eta^{5/2}(2P+1)} \left[ 1 + \frac{3}{4(2P+1)} \right] \tag{6.15}$$

Using these relations one can easily verify the validity of (6.6) even for other values of $n_r = 2, 3, \ldots$ Therefore, we come now to the problem by another way – assume that the Eq. (6.6) is valid and calculate $\langle r^{-1/2} \rangle$ with the aid of standard solution for considered potential [16], using the spectral formula

$$E = -\frac{m\alpha^2}{2\hbar^2 [1/2 + n_r + P]^2} \tag{6.16}$$

which for $2P = 1/2$ becomes

$$E = -\frac{8m\alpha^2}{\hbar^2 (4n_r + 3)^2} \tag{6.17}$$

Therefore, according to notation (6.4) we find

$$\eta = \frac{8m\alpha}{\hbar^2 (4n_r + 3)} \tag{6.18}$$

And finally

$$\langle r^{-1/2} \rangle = \frac{4\eta^{1/2}\Gamma(3/2+n_r)}{\pi n_r!(2n_r+3/2)}, \qquad n_r = 0,1,\cdots \tag{6.19}$$

Let us underline, that the relation (6.9) can have many other powerful applications, especially, together with virial like considerations. One more example is given below for singular oscillator.

- **Singular oscillator potential** has the form

$$V = -\frac{V_0}{r^2} + gr^2, \qquad V_0 > 0 \quad g > 0 \tag{6.20}$$

This potential is interesting for the Calogero model [38]. In this case the standard wave function of the Schrodinger equation is [39]

$$R = Cr^{-1/2+P} e^{-\frac{\sqrt{2mg}}{2\hbar}r^2} \left(\frac{2mg}{\hbar}\right)^{(P-1/2)/4} F\left(-n, 1+P; \sqrt{2mg/\hbar}\, r^2\right) \tag{6.21}$$

Energy levels is given by obvious relation

$$E = \hbar\omega(2n_r + 1 + P), \qquad \omega = 2\sqrt{\frac{2g}{m}}, \qquad n_r = 0,1,2,\ldots \tag{6.22}$$

Here the following notations are used

$$\sqrt{\frac{2m}{g}} E = 4(n+s)+3; \quad \xi = \frac{\sqrt{2mg}}{\hbar} r^2 \equiv \eta r^2; \quad s = \frac{1}{2}\left(-\frac{1}{2}+P\right) \tag{6.23}$$

According the definition (2.11)

$$a_{st} = C\left(\frac{2mg}{\hbar}\right)^{(P-1/2)/4} \tag{6.24}$$

Because

$$\langle r^s \rangle = \int_0^\infty R^2 r^{s+2} dr \tag{6.25}$$

we have to calculate

$$\langle r^s \rangle = \frac{a_{st}^2}{2\eta^{P+1+s/2}} \int_0^\infty e^{-\xi} \xi^{P+s/2} F^2(-n, 1+P; \xi) d\xi \tag{6.26}$$

Comparison to the Eq. (6.7) dictates

$$v = P+1+s/2; \quad k=1; \quad \gamma = P+1 \tag{6.27}$$

Therefore

$$\langle r^s \rangle = \frac{a_{st}^2}{2\eta^{P+1+s/2}} \frac{\Gamma(P+1)\Gamma(P+1+s/2)n!}{\Gamma(P+n+1)} \times$$
$$\left\{ 1 + \frac{n(s/2)(s/2+1)}{1^2(P+1)} + \frac{n(n-1)(s/2+2)(s/2+1)(s/2)(1-s/2)}{1^2 2^2 (P+1)(P+2)} + \ldots \right.$$
$$\left. + \frac{n!(-s/2-n)(-s/2+n-1)}{1^2 2^2 \ldots n^2 (P+1)(P+2)\ldots(P+n)} \right\} \tag{6.28}$$

This relation is analogous to (6.9). The normalization constant can be calculated inserting here $s=0$. We find

$$1 = \frac{a_{st}^2 \left[\Gamma(P+1)\right]^2 n!}{2\eta^{P+1} \Gamma(P+n+1)} \tag{6.29}$$

It follows from (6.28) and (6.24), that

$$C^2 = \frac{2\eta^{3/2}}{n_r!(2n_r+2P+1)} \frac{\Gamma(2P+1+n_r)}{\left[\Gamma(P+1)\right]^2} \tag{6.30}$$

One must verify the relation (5.13) for $\beta=2$, i.e.

$$a_{st}^2 = \frac{mE}{\hbar^2} \langle r^{-2} \rangle \tag{6.31}$$

According to (6.28) in case of $s=-2$ all terms in figural bracket disappear, except to first one and it follows (remember, that according to definition $2P = \beta/2+1$ we have $P=1$ for $\beta=2$):

$$\langle r^{-2} \rangle = \frac{a_{st}^2}{2\eta} \frac{\Gamma(1)\Gamma(2)n!}{\Gamma(2+n)} \tag{6.32}$$

Thus, (6.31) runs into

$$1 = \frac{mE}{\hbar^2} \frac{1}{2\eta(n+1)} \tag{6.33}$$

It follows

$$E = \frac{2\eta\hbar^2}{m}(n+1) \tag{6.34}$$

Inserting here $\eta$ from (6.23), we derive

$$E = 2\sqrt{\frac{2g}{m}}\hbar(n+1) \tag{6.35}$$

which coincides with proper energy (6.22) as $P=1$.

Therefore, we have finished the verification of virial relations in these cases also. It is remarkable to note that the more general relation (5.12) for singular oscillator may be used for arbitrary $P$ to relate various even degrees of $r$ to each other's.

## 7. CONCLUSIONS

Collecting all above derived results one can conclude that:

Modified hypervirial theorems (2.32)-(2.33) gave us true physical results in various considered cases. They coincide to the usual relations, when the additional terms are absent. In cases, when the additional contributions appear, modified relations give reasonable corrections and supplement deficient contributions. We have checked the validity of additional term for various potentials (regular and soft-singular) explicitly and established its legitimacy.

Equations (2.32) and (2.33) are sources for obtaining relations between mean values of various degrees of radius for large classes of potentials – both regular and soft-singular. Assigning some values to parameter $s$ and specified potentials, one can derive a generalization of Kramers' relation and correct the Ehrenfest theorem in a true direction, as well as higher order derivatives of radial function at the origin, which may have an application in the Van Royen-Waisscopf formula for decay probabilities. As a byproduct, some complicated integrals for hypergeometric functions are also derived, which are exhibited in the Appendix below.

**Appendix:**

Let consider various examples and find corresponding integrals:

2. *The Hulten potential:*

$$V = -\frac{V_0}{e^{\frac{r}{a}} - 1} \tag{A.1}$$

The solution of the Schrodinger equation in this case ($l=0$) is [36]

$$u = rR = C_{0n_r} e^{-\eta r} F\left(\alpha, \beta, \gamma; e^{-\frac{r}{a}}\right) \tag{A.2}$$

Here

$$\eta = \sqrt{-\frac{2mE_{0,n_r}}{\hbar^2}}, \qquad \alpha = \varepsilon + \sqrt{\varepsilon^2 + \lambda^2}, \qquad \beta = \varepsilon - \sqrt{\varepsilon^2 + \lambda^2}$$

$$\varepsilon = \eta a, \qquad \lambda = \left(\frac{2ma^2 V_0}{\hbar^2}\right)^{1/2} \tag{A.3}$$

$F$ is a Hypergeometric function, $C_{0n_r}$ - normalization constant, and energy levels are obtainable from the condition

$$F(\alpha, \beta, \gamma; 1) = 0 \tag{A.4}$$

For this potential Eq. (4.18) takes the form

$$\frac{C_1^2 \hbar^2}{2m} = \left\langle \frac{dV}{dr} \right\rangle = \frac{V_0}{a}\left(C_{0n_r}\right)^2 \int_0^\infty e^{-2\eta r} e^{\frac{r}{a}} \left(e^{\frac{r}{a}} - 1\right)^2 F^2\left(\alpha, \beta, \gamma; e^{-\frac{r}{a}}\right) dr \tag{A.5}$$

Behavior of (A.2) at the origin gives

$$\lim_{r \to 0} u(r) = -C_{0n_r} F'(\alpha, \beta, \gamma; 1) \frac{r}{a} \tag{A.6}$$

or

$$C_1 = -C_{0n_r} F'(\alpha, \beta, \gamma; 1) \frac{1}{a} \tag{A.7}$$

Then Eq. (4.18) gives

$$\int_0^\infty e^{-2\eta r} e^{\frac{r}{a}} \left( e^{\frac{r}{a}} - 1 \right)^2 F^2\left(\alpha, \beta, \gamma; e^{-\frac{r}{a}}\right) dr = -\frac{a\hbar^2}{2mV_0 a^2} F'^2(\alpha, \beta, \gamma; 1) \tag{A.8}$$

or, accounting notations (A.3), we have

$$\int_0^\infty e^{-2\eta r} e^{\frac{r}{a}} \left( e^{\frac{r}{a}} - 1 \right)^2 F^2\left(\alpha, \beta, \gamma; e^{-\frac{r}{a}}\right) dr = -\frac{a}{\lambda^2} F'^2(\alpha, \beta, \gamma; 1) \tag{A.9}$$

But from (A.4) and

$$2\eta = \frac{\alpha + \beta}{a} \tag{A.10}$$

and finally, we get

$$\int_0^\infty e^{-\frac{\alpha+\beta}{a} r} e^{\frac{r}{a}} \left( e^{\frac{r}{a}} - 1 \right)^2 F^2\left(\alpha, \beta, \gamma; e^{-\frac{r}{a}}\right) dr = -\frac{a}{\lambda^2} F'^2(\alpha, \beta, \gamma; 1) \tag{A.11}$$

Moreover, again from (A.3)

$$\lambda^2 = -\alpha\beta \tag{A.12}$$

Therefore

$$\int_0^\infty e^{-\frac{\alpha+\beta}{a} r} e^{\frac{r}{a}} \left( e^{\frac{r}{a}} - 1 \right)^2 F^2\left(\alpha, \beta, \gamma; e^{-\frac{r}{a}}\right) dr = \frac{a}{\alpha\beta} F'^2(\alpha, \beta, \gamma; 1) \tag{A.13}$$

Now we can use the recurrence relation for derivatives of this function [37]

$$F'(a, b, c; z) = \frac{ab}{c} F(a+1, b+1, c+1; z) \tag{A.14}$$

And at the end we have

$$\int_0^\infty e^{-\frac{\alpha+\beta}{a}r} e^{\frac{r}{a}} \left(e^{\frac{r}{a}}-1\right)^2 F^2\left(\alpha,\beta,\gamma; e^{-\frac{r}{a}}\right) dr = \frac{\alpha\beta a}{\gamma^2} F^2(\alpha+1,\beta+1,\gamma+1;1) \quad (A.15)$$

3. **The Morse potential:**

$$V(r) = D\left(e^{-2\alpha x} - 2e^{-\alpha x}\right); \qquad x = \frac{r-r_0}{r_0} \quad (A.16)$$

This potential has a wide application in Chemistry for studying two-atomic molecules.

The solution of the s-wave Schrodinger equation looks like [36]

$$u = rR = C_{0n_r} y^{\frac{\beta}{\alpha}} e^{-\frac{y}{2}} F(a,c,y) \quad (A.17)$$

Here $F$ is a confluent hypergeometric function, and the following notations are used

$$y = \frac{2\gamma}{\alpha} e^{-\alpha x}; \quad c = 2\frac{\beta}{\alpha}+1; \quad a = \frac{1}{2}c - \frac{\gamma}{\alpha}; \quad \beta^2 = -\frac{2mEr_0^2}{\hbar^2} > 0, \; \gamma^2 = \frac{2mDr_0^2}{\hbar^2} \quad (A.18)$$

Eigenvalues equation is

$$u(0) = F(a,c,y_0) = 0; \qquad y_0 = \frac{2\gamma}{\alpha} e^{\alpha} \quad (A.19)$$

Proceeding in a similar way as above, one finds the following equality

$$\int_0^\infty F^2\left(a,c,\frac{2\gamma}{\alpha}e^{-\alpha\left(\frac{r-r_0}{r_0}\right)}\right) \left(e^{-2\alpha\left(\frac{r-r_0}{r_0}\right)} - e^{-\alpha\left(\frac{r-r_0}{r_0}\right)}\right) dr = \frac{r_0}{2\gamma^2\alpha} y_0^{\frac{2\beta}{\alpha}+2} e^{-y_0} \frac{a^2}{c^2} [F(a+1,c+1;y_0)]^2 \quad (A.20)$$

Consider now relation (4.42), which gives

$$D\left\langle r\left(e^{-2\alpha\left(\frac{r-r_0}{r_0}\right)} - e^{-\alpha\left(\frac{r-r_0}{r_0}\right)}\right)\right\rangle = 2D\left\langle \left(e^{-2\alpha\left(\frac{r-r_0}{r_0}\right)} - e^{-\alpha\left(\frac{r-r_0}{r_0}\right)}\right)\right\rangle - 2E \quad (A.21)$$

In explicit form this equation means, according to notations (A.18)

$$\int_0^\infty F^2\left(a,c,\frac{2\gamma}{\alpha}e^{-\alpha\left(\frac{r-r_0}{r_0}\right)}\right)\left(e^{-2\alpha\left(\frac{r-r_0}{r_0}\right)} - e^{-\alpha\left(\frac{r-r_0}{r_0}\right)}\right) rdr = \frac{r_0}{\gamma^2\alpha} y_0^{c+1} e^{-y_0} \frac{a^2}{c^2}[F(a+1,c+1;y_0)]^2 + \frac{\alpha^2(c-1)^2}{4\gamma^2}$$

$$(A.22)$$

### 4. Wood-Saxon potential:

$$V(r) = -\frac{V_0}{1+e^{\frac{r-R}{a}}}; \qquad a << R \tag{A.23}$$

It is applied for description of neutron-nucleus interaction. For s-wave Schrodinger equation the solution is [36]

$$u = rR = C_{0n_r} y^{\nu}(1-y)^{\mu} F(\mu+\nu, \mu+\nu+1, 2\nu+1; y), \tag{A.24}$$

Here $F$ is a hypergeometric function and

$$y = \frac{1}{1+e^{\frac{r-R}{a}}}; \quad \nu = \beta; \quad \mu^2 = \beta^2 - \gamma^2; \quad \beta^2 = -\frac{2mEa^2}{\hbar^2} > 0, \quad \gamma^2 = \frac{2mV_0 a^2}{\hbar^2} \tag{4.25}$$

Spectrum is derived by the condition

$$u(0) = F(\mu+\nu, \mu+\nu+1, 2\nu+1; y_0) = 0; \qquad y_0 = \frac{1}{1+e^{-\frac{R}{a}}} \tag{A.26}$$

It is easy to show that

$$\lim_{r \to 0} u(r) = -C_{0n_r} y_0^{\nu+1}(1-y_0)^{\mu} F'(\mu+\nu, \mu+\nu+1, 2\nu+1; y_0) \frac{r}{a} \frac{e^{-\frac{R}{a}}}{1+e^{-\frac{R}{a}}} \tag{A.27}$$

So

$$C_1 = -C_{0n_r} y_0^{\nu+1}(1-y_0)^{\mu} F'(\mu+\nu, \mu+\nu+1, 2\nu+1; y_0) \frac{1}{a} \frac{e^{-\frac{R}{a}}}{1+e^{-\frac{R}{a}}} \tag{A.28}$$

Analogous consideration, as above, gives the following relation

$$\int_0^{\infty} F^2\left(\mu+\nu, \mu+\nu+1, 2\nu+1; \frac{1}{1+e^{\frac{r-R}{a}}}\right) \frac{e^{\frac{r-R}{a}}}{\left[1+e^{\frac{r-R}{a}}\right]^2} dr =$$

$$= -\frac{a^3}{\gamma^2}\left[y_0^{\nu+1}(1-y_0)^{\mu} F'(\mu+\nu, \mu+\nu+1, 2\nu+1; y_0)\frac{1}{a}\frac{e^{-\frac{R}{a}}}{1+e^{-\frac{R}{a}}}\right]^2$$

$$\tag{A.29}$$

We do not find these integrals (A.15), (A.20), (A.22) and (A.29) in accessible to us Tables and Handbooks.